\documentstyle[12pt,german]{article}
\def\i{{\rm i}}

\def\ybacu{YBa$_2$Cu$_3$O$_7$\ }

\newlength{\mycolwidth}%
\newlength{\myindents}%
\newenvironment{indented}[1]{%
\setlength{\mycolwidth}{#1}%
\setlength{\myindents}{\textwidth}
\addtolength{\myindents}{-\mycolwidth}
\begin{list}{}{\leftmargin0.5\myindents %
\rightmargin0.5\myindents \parsep\parskip}\item[]}{\end{list}}
\def\abstract{\if@twocolumn
\section*{\abstractname}
\else \renewcommand{\baselinestretch}{0.9}
\begin{center}
{\bf \abstractname\vspace{-1cm}\vspace{0pt}}
\end{center}
\begin{indented}{15cm}
\fi}
\def\endabstract{\if@twocolumn\else\end{indented}\vspace{0.3cm}\fi}
\advance\voffset by -2truecm
\advance\hoffset by -3.0 truecm
\advance\textheight by 3.5truecm
\renewcommand{\baselinestretch}{0.95}
\begin{document}
\selectlanguage{\USenglish}
\def\abstractname{\  }
\title{\renewcommand{\baselinestretch}{1.0}
\uppercase{Frequency Shifts and Linewidth Changes of Infrared-Active Phonons
in Double-Layered
High-Temperature Superconductors}}
\author{G.~Hastreiter, F.~Forsthofer, J.~Keller \\
{\rm Institut f"ur Theoretische Physik}\\
{\rm Universit"at Regensburg, Universit"atsstra"se 31, D-93040
Regensburg}\\{\rm Federal Republic of Germany}}
%
\date{(Received \hspace{5cm})}
\maketitle
\begin{abstract}
We calculate frequency shifts and changes in linewidths of infrared-active
phonons within a shell model for the bare phononic system coupled to an
electronic double-layer structure with inter-layer charge transfer. The
theoretical concept is applied to \ybacu\ yielding a good description of
experimental results in the normal state as well as at the transition to the
superconducting state.
\end{abstract}
\newpage
\begin{indented}{10cm}
\section{Introduction}
Raman- \cite{Thomsen,Friedl} and infrared-experi"-ments
\cite{Genzel} -- \cite{Bazhenov}
show a strong influence of the superconducting transition on phonon
frequencies and linewidths in \ybacu and similar compounds.
The experimental results on Raman-active  phonons are explai"-ned
by the theory  of Zeyher  and  Zwicknagl  \cite{ZeyZw.1,ZeyZw.2}.
Complementary  to their work the even greater variety  of effects
found   on  infrared-active   phonons,   e.\  g.\  narrowing   of
phonon-lines   \cite{Litvin}   and  softening  of  high-frequency
phonons \cite{Genzel}  -- \cite{Bazhenov}  in con\-nec\-tion with the
transition  into  the superconducting  state,  has been explained
qualitatively  \cite{Hast.1,Hast.3}  within a double-layer  model
for high-temperature superconductors.

In this paper  we link  the approaches  of refs.\  \cite{Hast.2},
where   a  mi\-cro\-scop\-i\-cal\-ly   based   model   for  a  Coulomb-type
elec\-tron-phonon interaction in layer systems was developed, with
the  calculation  of  the  electronic  susceptibility   for  the
double-layer  model from \cite{Hast.3}.   This results  in a {\it
microscopically    founded}   description   of   phonon
line width
\break changes  and frequency  shifts  which  fits to the experimental
data  in the normal  state  as well  as in the
superconducting state.

\section{Infrared absorption in  coupled phonon-layer systems}

We  are  going  to  investigate  the  infrared absorption  of  an
essentially  ionic system coupled to conduction  electrons  which
are confined  to two-dimensional  planes within a model presented
in  \cite{Hast.2,Hast.3}:    Bare  phonons  are  described  by  a
shell-model  of lattice dynamics,  bare electrons  are allowed to
move  almost  freely  within  the conducting  CuO$_2$-planes;  we
assume hopping in c-direction between different planes within one
unit cell but neglect transport  between different  double-layers
for simplicity.  We consider electron-phonon  coupling due to the
Coulomb-interaction   between   vibrating   ions  and  conduction
electrons.

In order  to determine  the response function  for IR absorption,
the dielectric  function, of a coupled electron-phonon  system it
is not sufficient to calculate a renormalized phononic Green's
function  and,  independently,   an  ---  also  renormalized  ---
electronic  susceptibility.    E\-lec\-tron-phonon   coupling  yields
``off-diagonal''  terms  in the  dielectric  function  which  may
change lineshapes of absorption lines.

This  can  be described  by a generalized  susceptibility  of the
coupled electron-phonon  system depending  on frequency  $\omega$
and   wavevector   ${\bf   q}$.   Its   inverse   is   given   by

\begin{equation}   \chi^{-1}_{\rm   gen}   ({\bf   q},\omega)   =
\left(\begin{array}{cc}  {\bf  G}^{-1}_{\rm  ph}  &  \gamma^+  \\
\gamma  & \underline\chi^{-1}  \end{array}  \right)  \:  \mbox{.}
\label{chigeninv} \end{equation}

Here, ${\bf G}_{\rm ph}$ is the bare phonon Green's function  which
is determined by

\begin{equation} \left[ {\bf 1} (\omega+\i\delta)^2  - {\bf
D}_0  \right]  M^{1/2}  {\bf  G}_{\rm   ph}  M^{1/2}  =  {\bf  1}
\label{PhonGF}  \end{equation}

with  the dynamical  Matrix  ${\bf  D}_0 ({\bf  q})$ and diagonal
matrices  $M$ containing  the vibrating  masses.   For reasons of
simplicity  we do not make any formal distinction  between  cores
and shells in the lattice dynamical shell-model  used for the
description  of the bare phonon system.  Instead,  we use shell
masses about $10^6$ times smaller than core masses. Thus, for a
system of $N$ ions ${\bf D}_0$ is a $(6N\times  6N)$ matrix whose
$3N$  lowest  eigenvalues  describe  the  dispersion  of the bare
vibrational system.

$\underline\chi  ({\bf q},\omega)$  is the susceptibility  of the
bare electronic system. In general this is a $(n\times n)$ matrix
where $n$ is the number  of conducting  layers per unit cell.  In
the case of a double-layer  system  and $\bf  q$ parallel  to the
planes  this matrix  can be transformed  into diagonal  form with
elements   describing    intra-   and   interband    excitations,
respectively \cite{Hast.3}.

Finally,  $\gamma({\bf  q})$ in (\ref{chigeninv})  is a $(n\times
6N)$ matrix containing  the electron-phonon  matrix elements.  We
calculate  these matrix elements according  to our ionic approach
described in detail in \cite{Hast.2}  (with ${\bf X}_{\mu\kappa}$
from \cite{Hast.2}  and ionic charge numbers ${\cal Z}_\kappa$ we
have $\gamma_{\mu\kappa}={\cal Z}_\kappa {\bf X}_{\mu\kappa}$ for
$\mu=1,2$, $\kappa = 1, \ldots , 6N$).

In order to calculate  the infrared  absorption  we still have to
multiply the matrix elements of $\chi_{\rm  gen}$ by the coupling
constants  to the  external  electric  field.   For the  case  of
infrared  active phonons with eigenvectors  perpendicular  to the
$a$-$b$-plane  and charge fluctuations  between the double layers
the dielectric function reads

\begin{equation}
\Delta\varepsilon(\omega)  = 4\pi A^+ \chi_{\rm gen}({\bf  q} \to
0, \omega ) A \label{epsilonir}\end{equation}

with
\begin{eqnarray}
A_{\kappa}&=&e{\cal Z}_\kappa \quad\mbox{f"ur}\quad \kappa=3\nu
\mbox{;\ } \nu=1,\ldots 2N \label{Airc1}\\
A_{6N+1}&=&\pm 0.5 e d \label{Airc2}\\
A_{6N+2}&=&\mp 0.5 e d \label{Airc3} \: \mbox{.}
\end{eqnarray}
All other components  of the external  field coupling vectors $A$
are zero. The electrical field couples to the polarization in the
crystal. For lattice vibrations this polarization is given by the
product  of ionic charge (or shell charge)  and the corresponding
displacement.   The latter is contained in $\chi_{\rm gen}$.  For
the electronic excitations the polarization  is determined by the
distance  $d$ between the two planes connected  by the electronic
hopping  process  and the charge transfered  between these layers
which depends  on the electronic  susceptibility.   The signs and
distances to be used in (\ref{Airc2}) and (\ref{Airc3}) depend on
the  way  the  inter-layer   hopping  process   is  taking  place
physically.

\section{Infrared absorption of \ybacu}

In the following
we apply  the theoretical  concepts  described  above  to \ybacu\
using  a  lattice-dynamical  shell  model  with  parameters  from
\cite{Kress}. As we are mainly interested in  general  trends  we
refrain from a ''fine-tun\-ing'' of the original shell-model parameters.

The  electronic  double-layer  system  is modelled  by two  bands
crossing  the Fermi surface  \cite{Hast.3}.   For simplicity  the
bands  are assumed  to have  zylindrical  symmetry  but an energy
distance  varying  from  zero  to  about  1000  cm$^{-1}$  in the
vicinity of the Fermi surface. The mean density of states is 0.85
eV$^{-1}$  for each band.  In the superconducting  state we use a
BCS-type approximation with equal gaps $2\Delta=320$ cm$^{-1}$ in
both bands.

The bare electronic susceptibility  is calculated numerically  as
described in \cite{Hast.3} taking into account vertex corrections
and screening effects in a random-phase approximation.  We assume
that  the  electronic   hopping   process  takes  place  via  the
copper-oxygen chains, while the yttrium layer is considered to be
completely  insulating.   Thus,  the distance  $d$ to be used  in
(\ref{Airc2}) and (\ref{Airc3}) is about 8.3 \AA.

The electron-phonon  coupling at large wavelengths  is completely
determined   by  the  Coulomb   interaction   between  ionic  and
electronic  charges without additional parameters as described in
\cite{Hast.2}.

Fig.\ 1a shows the result of the calculation for the {\it normal}
state.   The dashed  line represents  the shell model calculation
(with an artificial  broadening  of the phonon lines),  the solid
line represents  the results for the coupled system including the
electron-phonon  interaction.   Three prominent effects are to be
observed: {\it softening} of phonon frequencies, {\it broadening}
of  absorption  lines  and  strong  {\it  changes  of  oscillator
strengths} for different modes.

Most pronounced  are these effects  for the mode at 365 cm$^{-1}$
in the uncoupled  system,  where the yttrium-layer  moves against
the conducting CuO$_2$-planes \cite{Kress}. The frequency of this
mode is lowered by about 85 cm$^{-1}$  due to the electron-phonon
coupling,  the  linewidth  increases  as well  as the  oscillator
strength.   Actually,  such  a frequency  shift  for this mode is
observed experimentally going from YBa$_2$Cu$_3$O$_6$  to \ybacu\
\cite{Reichardt}.
Similar effects are found also for the modes at 154 cm$^{-1}$ and
508 cm$^{-1}$  (in the bare model) which both become softer by 13
cm$^{-1}$.

These  softenings  result  from the fact that ionic displacements
perpendicular  to the conducting  double  layer  induce  a charge
transfer  between  the planes of this double layer.   This charge
transfer  is accompanied  by a polarization  field  \cite{Hast.2}
which reduces the restoring forces onto the ion, the frequency of
the vibration becomes lower. The line broadening is caused by the
damping  of lattice  vibrations  due  to interband  particle-hole
excitations in the electronic system.

Somewhat less evident is the situation concerning  the changes in
the oscillator strength.  Here, one has to distinguish two cases:
firstly,  vibrations  of ions which are located  between  the two
planes connected  by the electronic  hopping  process, in our
case this would be e.\ g.\ the barium  ions; here the direction
of the electronic  polarization  due to the  hopping  process  is
antiparallel   to   the   polarization   resulting   from   ionic
displacements,  therefore  the  oscillator  strength  of the mode
(around 150 cm$^{-1}$)  decreases.   If on the other hand e.\ g.\
yttrium  ions  move  perpendicular  to the  planes,  the  induced
electronic polarization  is in the same direction as the phononic
polarization,   the  oscillator   strength  therefore  increases.
Experiments also show large oscillator strengths for barium modes
\cite{Genzel} in contradiction to our results. This may be due to
additional  charge transfer processes, e.\ g.\ in the CuO-chains,
which are not contained in our model.

Our numerical  results for
the oscillator strength depend to a large extent on the details of
the  electronic  band  structure,   while  the  results  for  the
frequency shifts are more independent of such details.
Our calculations  for the dielectric function also yield a direct
contribution   from  collective  electronic  charge  fluctuations
between  the layers.  It appears  at much higher frequencies  and
therefore is not shown in Fig.1.

What  happens  at  the  transition  to the  {\it  superconducting
state}?  The infrared absorption  spectrum in the superconducting
state is shown in Fig.\ 1b.
In  Figs.\ 2a and b we show the relative changes in frequency and
linewidth due to this transition  calculated within the framework
of our model.  The small symbols refer to different  experimental
results,   the   circles   are  the  results   of  our  numerical
calculations and fit well the experimental frequency shifts. {\it
All} phonons --- even those at high frequencies --- become softer
in the superconducting  state,  phonon  lines below 500 cm$^{-1}$
become narrower due to the loss of decay channels resulting  from
the  gap  formation  in  the  superconducting  state.  Above  500
cm$^{-1}$  damping increases  and phonon lines may become broader
in the superconducting state.  The strongest effects are observed
for the modes arround 300 cm$^{-1}$.

We have also calculated  phonon  spectra  at finite  wave vectors
parallel  to the planes.   Here  we find  that  the relative
frequency shifts and line width changes at the transition  to the
superconducting  state  become  much  smaller.   This  is confirmed by
recent neutron scattering experiments \cite{Reichardt}.

\section{Summary}
In this  work  we have  presented  calculations  for the infrared
absorption  of \ybacu using a microscopically  based double-layer
model for high-temperature  superconductors.

The  approximations  used  to describe  the  conduction  electron
system  are rather  crude in view of strong-cor\-re\-lation  effects.
Nevertheless, they allow for a treatment of realistic systems and
we regard  the results  to be qualitatively  correct  in the long
wavelength limit.

The electron-phonon coupling was described without any additional
parameters by the Coulomb interaction between the ions inside and
outside the CuO$_2$-planes  and the mobile electrons.   Again, we
believe that this type of interaction will dominate the interband
electron-phonon coupling in the limit of small wavevectors.

Our  model   calculations   explain   experimental   results   on
frequency  shifts  and  line  width  changes  of  infrared-active
c-axis phonons. In particular the softening of modes
at the transition to the superconducting  state is well described
by our model.  Concerning  the absolute postion of the absorption
peaks the agreement  with experimental  results could be improved
if we would adjust the parameters  of the
original  shell-model.   We did not do this, because  we think it
more  important  to first  improve  the  approximations  for  the
electronic  system.
In some experiments  \cite {Litvin} it
seems, that the softening  of phonon frequencies  sets in already
at  temperatures  above  the  superconducting   transition
temperature. At the moment it is not clear wether this effect can
be explained  by superconducting  fluctuations  or by some  other
mechanism like opening of a spin gap.

\vspace{1cm}

\noindent{\it Acknowledgement} ---
This work was supported by the Deutsche Forschungsgemeinschaft.

\newpage
\thispagestyle{empty}
\noindent{\bf Figure Captions}

\vspace*{2cm}

\noindent{\bf Fig.~1.}
Calculated  infrared  absorption  for \ybacu
without  electron-phonon   interaction  (dashed  line)  and  with
electron-phonon  interaction  (solid  line);  a)  results  in the
normal  state  ($T$=100K)  and  b) in the  superconducting  state
$T$=1K.

\medskip

\noindent{\bf Fig.~2.}
Frequency  shifts  (a) and line width
changes    (b)    of   infrared-active    c-axis    phonons    in
$R$Ba$_2$Cu$_3$O$_7$  ($R$ = rare earth).  The small symbols  are
experimental results from refs.~\cite{Genzel} -- \cite{Bazhenov}.
The circles  mark the frequency  shifts  and line  width  changes
calculated  in this work.

\end{indented}

\end{document}